\pdfoutput=1

\documentclass[aps,prd,preprint,tightenlines,superscriptaddress,
amsfonts,amssymb,amsmath,byrevtex,showpacs,nofootinbib]{revtex4-1}

\usepackage{amssymb,amsmath}
\usepackage{latexsym}
\usepackage{graphicx}
\usepackage{geometry}
\usepackage{epstopdf}
\usepackage{setspace}
\usepackage{hyperref}
\usepackage{amsthm}
\usepackage{amsmath}
\usepackage{amssymb}
\usepackage{pgfplots}
\usepackage{graphicx} 
\usepackage{mathrsfs}
\usepackage{subcaption}
\usepackage{color}
\usepackage{verbatim}
\usepackage{tensor}
\usepackage{xcolor}
\usepackage{braket}

\newcommand{\dd}{\textnormal{d}}

\newcommand{\ee}{\textnormal{e}}

\begin{document}
\title{Curvature invariants for the  Bianchi IX spacetime \\ filled with tilted dust }

\author{Nick Kwidzinski}
\email{nk@thp.uni-koeln.de}\affiliation{Institute for Theoretical Physics,
University of Cologne, Z\"{u}lpicher Strasse 77, 50937 K\"{o}ln, Germany}

\author{W{\l}odzimierz Piechocki}
\email{wlodzimierz.piechocki@ncbj.gov.pl}\affiliation{ Department
of Fundamental Research, National Centre for Nuclear Research,
Ho{\.z}a 69, 00-681 Warszawa, Poland}

\date{\today}
\begin{abstract}
We present an analysis of the Kretschmann and Weyl squared scalars for the general Bianchi IX model filled with tilted dust.
Particular attention is given to the asymptotic regime close to the singularity for which we provide heuristic considerations
supported by numerical simulations. The present paper is an extension of our earlier publication \cite{Kiefer:2018uyv}.
\end{abstract}


\maketitle

\section{Introduction}

Einstein's theory of general relativity, GR, suffers from singularities (see, e.g. \cite{Haw,HP,Seno,Uggla}).
These are pathologies of spacetime, which determine the limits of the validity of GR. They also signal the occurrence of interesting phenomena
like black holes and the cosmological big bang, which seem to be  supported by observational data.

The present paper is an extension of our recently published paper \cite{Kiefer:2018uyv} on the evolution towards the singularity of the {\it general} Bianchi IX
spacetime. The dynamics of this model underlies the dynamics of the Belinski-Khalatnikov-Lifshitz (BKL) scenario (for an overview see e.g. \cite{Bel}),
which is conjectured to describe the approach towards a  {\it generic}  spacelike singularity of GR. As far as we know, the time evolution of
particular curvature invariants of the general BIX has not been considered yet.

The commonly known singularity theorems (see e.g. \cite{Haw,HP,Seno}) predict the existence of incomplete geodesics, but say little about other features
of singularities. Studying particular curvature invariants gives more characteristics of these pathologies.
The blowing up of the Kretschmann scalar was proved rigorously by Ringstrom \cite{Ringstrom} in the vacuum case of the BIX model (mixmaster universe).
Barrow and Hervik \cite{Barrow_Hervik_2002} have studied the Weyl tensor and provided asymptotic expression for homogeneous cosmological models
filled with non-tilted perfect fluids.   The behaviour of the Weyl squared scalar is of particular interest since it is conjectured that this invariant
acts as a measure of gravitational entropy \cite{Penrose_Weyl,Barrow_Hervik_2002}.

In our paper we examine the time evolution of the Kretschmann scalar for the non-diagonal  (general) BIX spacetime which requires
the coupling, for instance,  to a {\em tilted} matter field. The latter is chosen to be dust (for simplicity). The difference in the dynamics between
the diagonal and non-diagonal cases has been considered in \cite{Czuchry:2014hxa}.

Our paper is organized as follows:
We first review in section \ref{sec:general_BIX} the Hamiltonian formulation of the dust filled Bianchi IX model  which was employed for the numerical
simulation of the dynamics in \cite{Kiefer:2018uyv}. In section \ref{sec:Kretschmann} we compute an expression for the Kretschmann scalar which allows
for a numerical evaluation via the  framework based on the Hamiltonian formulation.

Section \ref{sec:asymptotic_regime} is devoted to the asymptotic regime close to the singularity. We provide heuristic considerations predicting
the behaviour of the Kretschmann scalar and support our result with a numerical analysis. We conclude in Sec. \!\ref{con}.

\section{Hamiltonian formulation of the dust filled Bianchi IX model}
\label{sec:general_BIX}
An examination of spatially homogeneous models with tilted fluids is given in \cite{King_Ellis_1973}.
The Bianchi IX model with  moving matter had been studied in \cite{Matzner_Shepley_Warren_1970,Grishchuk_Doroshkevich_Lukash_1972}.
The Hamiltonian formulation of the Bianchi IX model filled with tilted dust was first derived in a series of papers by Ryan \cite{Ryan_1971a,Ryan_1971b}.  For a similar formulation concerning the other Bianchi models see, for example, \cite{Ryan_HC,Ryan_Shepley,Jantzen:2001me}.
The metric in the ADM form is given by
\begin{equation}
\begin{aligned}
&\dd s^2=- N^2 \dd t^2 +
   h_{ij}\left(N^i\dd t+\sigma^i \right)\otimes\left(N^j\dd t+\sigma^j \right).
\end{aligned}
\label{eq:Non_diagonal_BIX_metric}
\end{equation}
The basis one-forms\footnote{Throughout this work we choose the units such that $\frac{3}{4\pi G}\int \sigma^1\wedge\sigma^2\wedge \sigma^3=1$, which correspond the setting $\kappa={8\pi G}/{c^4}=6$.}
 satisfy the relation
\begin{equation}
\dd \sigma^i=-\frac{1}{2}C^i_{jk}\sigma^j \wedge \sigma^k \ ,
\end{equation}
with $C^i_{jk}=\varepsilon_{ijk}$ being the structure constants of the Lie algebra $\mathfrak{so}(3, \mathbb{R})$.
We parametrize the metric coefficients  in this frame as follows:
\begin{equation}
h_{ij} =  O_i{}^{k}  O_j{}^{l} \bar{h}_{ k l }\, ,
\end{equation}
where
\begin{equation}
\bar{h}\equiv\left\{
\bar{h}_{i j}
\right\} =
 \ee^{2\alpha}\text{diag}\left(
\ee^{2\beta_+ + 2\sqrt{3}\beta_-},
\ee^{2\beta_+ - 2\sqrt{3}\beta_-},
\ee^{-4\beta_+ }
\right)\equiv\text{diag}\left(\Gamma_1 ,\Gamma_2 , \Gamma_3 \right) .
\end{equation}
The variables $\alpha$, $\beta_+$, and $\beta_-$ are known as
Misner variables.\footnote{Note that Misner originally used the variable $\Omega=-\alpha$.}
 The scale factor $\exp(\alpha)$ is related to the volume,
while the anisotropy factors $\beta_+$ and $\beta_-$ describe the shape
of the universe.
The variables $\Gamma_1$, $\Gamma_2$  and  $\Gamma_3 $ were used by
BKL in their original analysis \cite{bkl}.
In addition, we have  introduced the $SO(3,\mathbb{R})$ matrix $O\equiv\left\{O_i{}^{j}\right\}\equiv
O_{\theta}O_{\phi}O_{\psi}$,
which will be parameterized by a set of Euler angles,
$\left\{ \theta,\phi,\psi \right\}\in
[0,\pi]\times[0,2\pi]\times [0,\pi]$. Explicitly,
\begin{equation}
\label{matrixO}
\begin{aligned}
O_{\psi}=\left(
\begin{array}{ccc}
\cos ( \psi ) & \sin ( \psi ) & 0 \\
- \sin ( \psi ) & \cos ( \psi ) & 0 \\
0 & 0 & 1
\end{array}
\right) \ , \quad
O_{\theta}=\left(
\begin{array}{ccc}
1 & 0 & 0 \\
0 &\cos ( \theta ) & \sin ( \theta )  \\
0 &- \sin ( \theta ) & \cos ( \theta )
\end{array}
\right)\, ,
\\
O_{\phi}=\left(
\begin{array}{ccc}
\cos ( \phi ) & \sin ( \phi ) & 0 \\
- \sin ( \phi ) & \cos ( \phi ) & 0  \\
0 & 0 & 1
\end{array}
\right)\, .\qquad\qquad\quad\qquad
\end{aligned}
\end{equation}
The Euler angles $\theta$, $\phi$, and $\psi$ are
dynamical quantities and describe nutation,
precession, and pure rotation of the principal axes, respectively.
The Hamiltonian and momentum constraints are given by
\begin{equation}
\begin{aligned}
&\mathcal{H} + \mathcal{H}^{(m)}
=
\frac{\ee^{-3\alpha}}{2}\left(-p_\alpha^2 +p_+^2 +p_-^2
+\frac{l_1^2}{I_1}+\frac{l_2^2}{I_2}+\frac{l_3^2}{I_3}
- \frac{\ee^{6\alpha}}{6} {}^{(3)}R
+ 2\ee^{3\alpha} p_T \sqrt{1+h^{ij}u_i u_j}
\right),
\\
&\mathcal{H}_i  +\mathcal{H}^{(m)}_i
= O_i{}^j\left( l_j - C p_T v_j \right)  \ .
\end{aligned}
\label{eq:Hamiltonian+diffeo_constraints_dust}
\end{equation}
The $l_i$ denote (non-canonical) angular momentum-like variables:
\begin{equation}
 l_1 \equiv\frac{\ee^{3\alpha}}{N}I_1 \omega^{2}{}_3  , \quad
 l_2 \equiv\frac{\ee^{3\alpha}}{N}I_2 \omega^{1}{}_3    \quad \text{and} \quad
 l_3 \equiv\frac{\ee^{3\alpha}}{N}I_3 \omega^{2}{}_1  \ ,
\end{equation}
where we defined the `angular velocities'
$
\boldsymbol{\omega}=\left\{\omega^{i}{}_{j} \right\} \equiv O^{T}\dot{O}
$ and the `moments of inertia'
\begin{equation}
3 I_1  \equiv\sinh^2 \left( 3\beta_+ - \sqrt{3}\beta_-  \right)  , \quad
3 I_2  \equiv \sinh^2 \left( 3\beta_+ + \sqrt{3}\beta_-  \right)  , \quad
3 I_3  \equiv \sinh^2 \left( 2\sqrt{3}\beta_- \right) .
\end{equation}
The $l_i$ obey the Poisson bracket algebra
$
\left\{ l_i,l_j \right\} =-C^k_{ij}{}l_k
$.
The variable $p_T$, which is in fact a constant of motion, denotes the momentum canonically conjugate to the `dust time' $T$, that is, the proper time measured by observes co-moving with the dust particles.
The three-curvature scalar on spatial hypersurfaces of constant
coordinate time $t$ is given by
\begin{equation}
{}^{(3)} R =  -\frac{\ee^{-2\alpha}}{2}
 \left(
\ee^{-8\beta_+}-4\ee^{-2\beta_+} \cosh \left(2\sqrt{3}\beta_- \right)
+ 2\ee^{4\beta_+}
\left[ \cosh\left(
4\sqrt{3}\beta_-
\right)
-1
\right]
\right) .
\label{eq:3R_BIX}
\end{equation}
The formalism described so far is not entirely canonical
and must be complemented by the geodesic equation for the dust particles
\begin{equation}
\left( \partial_t+\boldsymbol{\omega} \right)\vec{v} \ =\frac{ N C  \left[ \vec{v}\times ({{\bar{h}}}^{-1}\vec{v} )\right]}{\sqrt{1+C^2  \vec{v}^T {{\bar{h}}}^{-1}\vec{v} }}
\label{eq:v_geodesic}  \ ,
\end{equation}
where
$C O \vec{v}\equiv  \left(u_1 ,\ u_2, u_3 \right)^T$
, $
C^2\equiv(u_1)^2+(u_2)^2+(u_3)^2
$ is a constant of motion and ``$\times$'' denotes the usual cross product in 3 dimensional Euclidean space.
The angular velocities $\boldsymbol{\omega}$ can be eliminated from the equations of motion by using the momentum constraints.
The equations of motion have a lengthy form and were given in \cite{Kiefer:2018uyv}.
Moreover, the Hamiltonian formulation can be employed to obtain a qualitative picture of the dynamics (see e.g. \cite{Kiefer:2018uyv,bkl,Ryan_1971a,Ryan_1971b,Jantzen:2001me}).

Our numerical simulations were performed by using the gauge $N=\ee^{3\alpha}$ and $N^i=0$. Setting $N=\ee^{3\alpha}$ ``moves'' the singularities to $t=\pm \infty$. This is required  to resolve the oscillations when evolving the system towards the singularity. Using the shooting method described in \cite{Kiefer:2018uyv} we numerically obtain the solution in terms of the variables
\begin{equation}
\log \Gamma_i , \ (\log \Gamma_i)^{\cdot} , \ v_i , \quad i=1,2,3
\label{eq:variables}
\end{equation}
as functions of $t$ over some finite time interval $[t_1, t_2]$. We restrict our attention to the so called tumbling case, that is, all $v_i$ are non-zero initially.
This case might be considered as the most generic one in the context of the model under consideration. For convenience, we restrict ourselves in this work to the
numerical solution which was already considered in \cite{Kiefer:2018uyv}.
 The part of the solution plotted in Fig. \ref{fig:bkl} extends over one Kasner era. We  regard this  solution to occur at the transition
into the asymptotic regime.

The numerical accuracy of our solution has been confirmed by checking the (approximate) preservation of the Hamiltonian constraint $\mathcal{H}=0$,
and  the constant of motion $C^2-1=0$. Both $\mathcal{H}$ and $C^2-1$ stay at order $10^{-15}$. We believe that the effect of chaos is negligible in a single
Kasner era. For further details see  \cite{Kiefer:2018uyv}.

\section{Calculation of the Kretschmann scalar}
\label{sec:Kretschmann}
We are interested in the temporal evolution of curvature invariants  when approaching the singularity.
Particular interest lies in the evolution of the Kretschmann scalar, which can be decomposed as
\begin{equation}
R_{\mu\nu\lambda\sigma}R^{\mu\nu\lambda\sigma}=C_{\mu\nu\lambda\sigma}C^{\mu\nu\lambda\sigma}+2R_{\mu \nu }R^{\mu \nu }-\frac{1}{3}R^2 \ ,
\label{eq:Kretschmann_decomposition}
\end{equation}
where $C^{\mu}{}_{\nu\lambda\sigma}$ is the Weyl tensor, $R_{\mu \nu }$ is the Ricci tensor and $R$ the Ricci scalar.
For our purposes it is convenient to make use of the constraints and the equations of motion to simplify the expressions such that they are suited for a numerical evaluation.   We will do so throughout the calculation in this section and bring our expression into a form that is ready for a numerical evaluation.
This means that all expressions should only involve the variables \eqref{eq:variables} as well as
the constants $p_T'\equiv 12p_T$ and $C$.
Furthermore we shall use the quasi-Gaussian gauge $N^i=0$ while keeping the lapse $N$ unspecified. We now proceed by calculating the terms on the right-hand side of equation \eqref{eq:Kretschmann_decomposition}.

 From the Einstein field equations $R_{\mu\nu}-\frac{1}{2}Rg_{\mu\nu}=\kappa T_{\mu\nu}=\kappa\rho u_\mu u_\nu $
  it follows that we can write
\begin{align}
R_{\mu \nu }R^{\mu \nu }  = \kappa^2 T_{\mu \nu }T^{\mu \nu }
=\kappa^2\rho^2 \quad \text{and} \quad
R  = -\kappa T^{\mu}{}_{\mu} =\kappa \rho
 \ .
\end{align}
Recall that in the model under consideration
\begin{equation}
\rho=\frac{p_T \ee^{-3\alpha}}{\sqrt{1+ h^{ij} u_i u_j}}=\frac{p_T }{\sqrt{\Gamma_1\Gamma_2 \Gamma_3+ C^2 \left(\Gamma_2\Gamma_3 v_1^2+\Gamma_1\Gamma_3v_2^2+\Gamma_1\Gamma_2 v_3^2\right)}} \ .
\end{equation}
 We conclude that the Ricci part $2R_{\mu \nu }R^{\mu \nu }-\frac{1}{3}R^2$ of the Kretschmann scalar blows up as \begin{equation}
 \left[
\Gamma_1\Gamma_2 \Gamma_3+ C^2 \left(\Gamma_2\Gamma_3 v_1^2+\Gamma_1\Gamma_3v_2^2+\Gamma_1\Gamma_2 v_3^2\right)\right]^{-1}
 \end{equation}
 when approaching the singularity.
The calculation of the Weyl part of the Kretschmann scalar, however, is less trivial.
 The 3+1 split allows for a decomposition of the Weyl tensor into electric and magnetic part (see e.g. \cite{Alcubierre}) according to
\begin{equation}
\begin{aligned}
& C_{\mu\nu\lambda\sigma}C^{\mu\nu\lambda\sigma} =
8\left(
E_{ij}E^{ij}-B_{ij}B^{ij}
\right) \quad \text{where}
\\
& E_{ij}  =  K_{ij}K_k{}^k - K_i{}^k K_{jk}+ {}^{(3)}R_{ij}
 - \frac{\kappa}{2} \left[S_{ij} + \frac{1}{3}h_{ij}\left( 4\epsilon - S^i{}_{i}\right) \right]
   \ ,
\\
& B_{ij} =   \epsilon_{ikl}\left[ D^k K_{j}{}^l - \frac{\kappa}{2} \delta^k_j j^l \right] \ ,
\end{aligned}
\label{eq:Weyl_EB}
\end{equation}
with $\epsilon_{ikl}$ being the Levi-Civita tensor. The other objects involved in the decomposition are explained below.
Now let  $P^\mu_\nu = \delta^\mu_\nu+ n^{\mu}n_{\nu}$ denote the projector  onto spatial hypersurfaces orthogonal to the normal vector $\{n^{\mu} \}=(1/N,0,0,0)$, that is $P^\mu_i = \delta^\mu_i$, $P^\mu_0=0$.  $D_i$ denotes the 3 dimensional covariant derivative on these hypersurfaces.
The quantities involved in equation (\ref{eq:Weyl_EB}) are
\begin{equation}
\begin{gathered}
 \epsilon = n^\mu n^\nu T_{\mu \nu} = \rho (1+u^i u_i) \ ,
\\
 S_{ij}= P^\mu_i P^\nu_jT_{\mu \nu}=\rho   u_i u_j \ ,
\\
 j^i=-P^{i\mu} n^\nu T_{\mu \nu} = p_T u^i/\sqrt{h} \ ,
\end{gathered}
\end{equation}
where $\epsilon$, $S_{ij}$ and $j^i$ are the energy density, the shear density and the momentum density as measured by Eulerian observers (observers with four velocity $n^{\mu} $).

A  direct calculation yields
\begin{equation}
\begin{aligned}
4N^2\Gamma_1\Gamma_2\Gamma_3B_{ij}B^{ij} & = \left( \Gamma_1^2+\Gamma_2^2+\Gamma_3^2\right)
\left[ \frac{\dot{\Gamma}_1}{\Gamma_1}+\frac{\dot{\Gamma}_2}{\Gamma_2}+\frac{\dot{\Gamma}_3}{\Gamma_3} \right]^2
\\
 & +
 \left( \Gamma_1+\Gamma_2+\Gamma_3\right)
 \left( \dot{\Gamma}_1+\dot{\Gamma}_2+\dot{\Gamma}_3\right)
 \left( \frac{\dot{\Gamma}_1}{\Gamma_1}+\frac{\dot{\Gamma}_2}{\Gamma_2}+\frac{\dot{\Gamma}_3}{\Gamma_3} \right)
  \\
 &+
 \frac{1}{4}\left( \Gamma_1+\Gamma_2+\Gamma_3\right)^2\left[
  \frac{\dot{\Gamma}_1^2}{\Gamma_1^2}+\frac{\dot{\Gamma}_2^2}{\Gamma_2^2}+\frac{\dot{\Gamma}_3^2}{\Gamma_3^2}
-
3\left( \frac{\dot{\Gamma}_1}{\Gamma_1}+\frac{\dot{\Gamma}_2}{\Gamma_2}+\frac{\dot{\Gamma}_3}{\Gamma_3} \right)^2
\right]
\\
&
+\frac{N^2C^2 p_T'^2\left( \Gamma_1+\Gamma_2+\Gamma_3\right)^2}{24\Gamma_1\Gamma_2\Gamma_3}\left [
\frac{v_1^2}{I_1}+\frac{v_2^2}{I_2}+\frac{v_3^2}{I_3}
\right] \ .
\end{aligned}
\end{equation}

Let us now turn to the computation of $E_{ij}E^{ij}$, which can be written out as
\begin{equation}
\begin{aligned}
E_{ij}E^{ij}
= &
K^i{}_j   K^j{}_k \left(  K^k{}_l  K^l{}_i
- 2 K^k{}_i K^l{}_l \right)
+ K^i{}_j   K^j{}_i  ( K^l{}_l )^2
\\
& - 2 \left( K^i{}_l K^{lj}-K^l{}_l K^{ij}  \right) \ {}^{(3)}R_{ij}
+ \
{}^{(3)}R_{ij}{}^{(3)}R^{ij}
\\
& +
6 \rho \left(K^i{}_lK^{lj} -K^l{}_l K^{ij} - {}^{(3)}R^{ij} \right)u_i u_j
\\
& +\rho\left( 8 + 6 u_k u^k   \right) \left[K^i{}_jK^{j}{}_{i}-  (K^l{}_l)^2 -{}^{(3)}R  \right]
\\
& + \rho^2 \left[ 54\left( u_k u^k  \right)^2 + 96\ u_k u^k  +48\right] \ .
\end{aligned}
\label{eq:E_squared}
\end{equation}
We now evaluate the single terms.
 The term in the first line of \eqref{eq:E_squared} right after the equal sign can be written as
\begin{equation}
\begin{aligned}
&8N^4 \left[  K^i{}_j   K^j{}_k \left(  K^k{}_l  K^l{}_i
- 2 K^k{}_i K^l{}_l \right)
+ K^i{}_j   K^j{}_i  ( K^l{}_l )^2 \right] =
\\
&\left[ (\log \Gamma_1)^\cdot(\log \Gamma_2)^\cdot\right]^2
+\left[ (\log \Gamma_1)^\cdot(\log \Gamma_3)^\cdot\right]^2
+\left[ (\log \Gamma_2)^\cdot(\log \Gamma_3)^\cdot\right]^2
\\
&+(\log \Gamma_1)^\cdot(\log \Gamma_2)^\cdot(\log \Gamma_3)^\cdot
\left[(\log \Gamma_1)^\cdot + (\log \Gamma_2)^\cdot + (\log \Gamma_3)^\cdot \right]
\\
&+N^4 C^4 p_T'^{\ 4} \left[
\frac{v_1^4}{\Gamma_1^2 (\Gamma_2-\Gamma_3)^4}
+\frac{v_2^4}{\Gamma_2^2 (\Gamma_1-\Gamma_3)^4}
+\frac{v_3^4}{\Gamma_3^2 (\Gamma_1-\Gamma_2)^4}
\right.
\\
&\left.
+
\frac{2v_1^2 v_2^2}{\Gamma_1 \Gamma_2\left(\Gamma_1-\Gamma_3 \right)^2\left(\Gamma_2-\Gamma_3 \right)^2}
\right.
\\
&\left.
+
\frac{2v_2^2 v_3^2}{\Gamma_3 \Gamma_2\left(\Gamma_2-\Gamma_1 \right)^2\left(\Gamma_3-\Gamma_1 \right)^2}
+
\frac{2v_1^2 v_3^2}{\Gamma_1 \Gamma_3\left(\Gamma_1-\Gamma_2 \right)^2\left(\Gamma_3-\Gamma_2 \right)^2}
\right]
\\
&
-\frac{N^2 C^2 p_T'^{\ 2}v_1^2}
{\Gamma_1 (\Gamma_2-\Gamma_3)^2}
\left[
(\log \Gamma_1)^\cdot
\left[
 (\log \Gamma_2)^\cdot+(\log \Gamma_3)^\cdot-(\log \Gamma_1)^\cdot\right]
+2 (\log \Gamma_2)^\cdot (\log \Gamma_3)^\cdot
\right]
\\
&
-\frac{N^2 C^2 p_T'^{\ 2}v_2^2}
{\Gamma_2 (\Gamma_1-\Gamma_3)^2}
\left[
(\log \Gamma_2)^\cdot
\left[
 (\log \Gamma_1)^\cdot+(\log \Gamma_3)^\cdot-(\log \Gamma_2)^\cdot\right]
+2 (\log \Gamma_1)^\cdot (\log \Gamma_3)^\cdot
\right]
\\
&
-\frac{N^2 C^2 p_T'^{\ 2}v_3^2}
{\Gamma_3 (\Gamma_2-\Gamma_1)^2}
\left[
(\log \Gamma_3)^\cdot
\left[
 (\log \Gamma_2)^\cdot+(\log \Gamma_1)^\cdot-(\log \Gamma_3)^\cdot\right]
+2 (\log \Gamma_2)^\cdot (\log \Gamma_1)^\cdot
\right] \ .
\end{aligned}
\label{eq:E_squared_term1}
\end{equation}
We denote the three-dimensional Ricci tensor of the diagonal model by ${}^{(3)}\bar{R}_{ij}$. The three-dimensional Ricci tensor of the non-diagonal model can then obtained via rotation according to
${}^{(3)}R_{ij}=O_i{}^k O_j{}^l \ {}^{(3)}\bar{R}_{kl}$. The only non-vanishing components of ${}^{(3)}\bar{R}_{ij}$ are given by
\begin{equation}
\begin{aligned}
{}^{(3)}\bar{R}_{11} &=  1 + \frac{\Gamma_1^2}{2\Gamma_2 \Gamma_3} -\frac{\Gamma_2}{2\Gamma_3}-\frac{\Gamma_3}{2\Gamma_2}
\\
{}^{(3)}\bar{R}_{22} &=  1 + \frac{\Gamma_2^2}{2\Gamma_1 \Gamma_3} -\frac{\Gamma_1}{2\Gamma_3}-\frac{\Gamma_3}{2\Gamma_1}
\\
{}^{(3)}\bar{R}_{33} &=  1 + \frac{\Gamma_3^2}{2\Gamma_1 \Gamma_2} -\frac{\Gamma_1}{2\Gamma_2}-\frac{\Gamma_2}{2\Gamma_1}
\ .
\end{aligned}
\end{equation}
The first term in the second line of \eqref{eq:E_squared} reads
\begin{equation}
\begin{aligned}
&-
 2 \left( K^i{}_l K^{lj}-K^l{}_l K^{ij}  \right) \ {}^{(3)}R_{ij}=
\frac{\left(\Gamma_2+\Gamma_3-\Gamma_1\right)
(\log \Gamma_2)^\cdot
(\log \Gamma_3)^\cdot}{2N^2\Gamma_2\Gamma_3}
\\
& \quad +
\frac{\left(\Gamma_1+\Gamma_3-\Gamma_2\right)
(\log \Gamma_1)^\cdot
(\log \Gamma_3)^\cdot}{2N^2\Gamma_1\Gamma_3}
+
\frac{\left(\Gamma_1+\Gamma_2-\Gamma_3\right)
(\log \Gamma_1)^\cdot
(\log \Gamma_2)^\cdot}{2N^2\Gamma_2\Gamma_1}
\\
& \quad
+
\frac{C^2 p_T'^{\ 2}}{2\Gamma_1\Gamma_2\Gamma_3}\left[
\frac{\left(\Gamma_1-\Gamma_2-\Gamma_3 \right)v_1^{\ 2}}{\left(\Gamma_2-\Gamma_3 \right)^2}
+
\frac{\left(\Gamma_2-\Gamma_1-\Gamma_3 \right)v_2^{\ 2}}{\left(\Gamma_1-\Gamma_3 \right)^2}
+
\frac{\left(\Gamma_3-\Gamma_1-\Gamma_2 \right)v_3^{\ 2}}{\left(\Gamma_1-\Gamma_2 \right)^2}
\right] \ .
\end{aligned}
\end{equation}
The three-dimensional Ricci squared scalar can be written as
\begin{equation}
\begin{aligned}
{}^{(3)}R_{ij}{}^{(3)}R^{ij} =
\frac{
\left(\Gamma_1^2  -12I_1 \Gamma_2 \Gamma_3 \right)^2
+\left(\Gamma_2^2  -12I_2 \Gamma_1 \Gamma_3 \right)^2
+\left(\Gamma_3^2  -12I_3 \Gamma_1 \Gamma_2 \right)^2
}{ 4\left(\Gamma_1 \Gamma_2 \Gamma_3 \right)^2}
\ .
\end{aligned}
\end{equation}
The term in the third line of \eqref{eq:E_squared} becomes
\begin{equation}
\begin{aligned}
& 6 \rho \left(K^i{}_lK^{lj} -K^l{}_l K^{ij} - {}^{(3)}R^{ij} \right)u_i u_j =
\\
&
\frac{3 \rho C^3 p_T' v_1 v_2 v_3}{N\sqrt{\Gamma_1 \Gamma_2 \Gamma_3 }}
\left[
\frac{(\log \Gamma_1)^\cdot}{\Gamma_3-\Gamma_2}
+  \frac{(\log \Gamma_2)^\cdot}{\Gamma_1-\Gamma_3}
+  \frac{(\log \Gamma_3)^\cdot}{\Gamma_2-\Gamma_1}
\right]
\\
&+\frac{3\rho C^4 p_T'^{\ 2}}{2}\left[
\frac{(\Gamma_2-\Gamma_3)^2 v_2 v_3}
{\Gamma_2\Gamma_3(\Gamma_1-\Gamma_2)^2(\Gamma_1-\Gamma_3)^2}
\right.
\\
&+
\left.
\frac{(\Gamma_1-\Gamma_3)^2 v_1 v_3}
{\Gamma_1\Gamma_3(\Gamma_2-\Gamma_3)^2(\Gamma_2-\Gamma_1)^2}
+\frac{(\Gamma_1-\Gamma_2)^2 v_1 v_2}
{\Gamma_1\Gamma_2(\Gamma_3-\Gamma_1)^2(\Gamma_3-\Gamma_2)^2}
\right]
\\
&+
\frac{3\rho C^2}{\Gamma_1 \Gamma_2 \Gamma_3 }
\left[ \left(
(\Gamma_2-\Gamma_3)^2 -\Gamma_1^2
- \frac{\Gamma_1\Gamma_2\Gamma_3}{2N^2}
(\log \Gamma_1)^\cdot \left[
(\log \Gamma_2)^\cdot+(\log \Gamma_3)^\cdot
\right]
\right)\frac{v_1^2}{\Gamma_1}
\right.&
\\
&\left.
 +
\left(
(\Gamma_1-\Gamma_3)^2 -\Gamma_2^2
- \frac{\Gamma_1\Gamma_2\Gamma_3}{2N^2}
(\log \Gamma_2)^\cdot \left[
(\log \Gamma_1)^\cdot+(\log \Gamma_3)^\cdot
\right]
\right)\frac{v_2^2}{\Gamma_2}
\right. &
\\
&\left.
 +
\left(
(\Gamma_1-\Gamma_2)^2 -\Gamma_3^2
- \frac{\Gamma_1\Gamma_2\Gamma_3}{2N^2}
(\log \Gamma_3)^\cdot \left[
(\log \Gamma_1)^\cdot+(\log \Gamma_2)^\cdot
\right]
\right)\frac{v_3^2}{\Gamma_3}
\right] &
\end{aligned}
\end{equation}
We can use the Hamiltonian constraint to simplify
\begin{equation}
K^i{}_jK^{j}{}_{i}-  (K^l{}_l)^2 -{}^{(3)}R = - p_T'
\sqrt{\frac{1+u_iu^i}{\Gamma_1\Gamma_2\Gamma_3}}\ .
\end{equation}
We therefore obtain a simple expression for the term in the fourth line of \eqref{eq:E_squared}.
Since we have direct numerical access to the quantities  in the fourth and fifth line of \eqref{eq:E_squared}, we will not manipulate them further.

It is well known that the Weyl squared scalar vanishes for the Friedmann models. The dust filled closed Friedmann universe is included in the model under consideration as the particular case for which $\Gamma_1=\Gamma_2=\Gamma_3$ and $C=0$.
As a consistency check of our calculation we convinced ourselves that the Weyl squared scalar vanishes for these restrictions. We find that
$B_{ij}B^{ij}$ and $E_{ij}E^{ij}$ vanish separately and hence $C_{\mu\nu\lambda\sigma}C^{\mu\nu\lambda\sigma}=0$ as expected.

\begin{figure}[!ht]
\centering
\hspace{-2.5cm}
	\begin{subfigure}[t]{0.4\textwidth}
		\includegraphics[width=8cm]{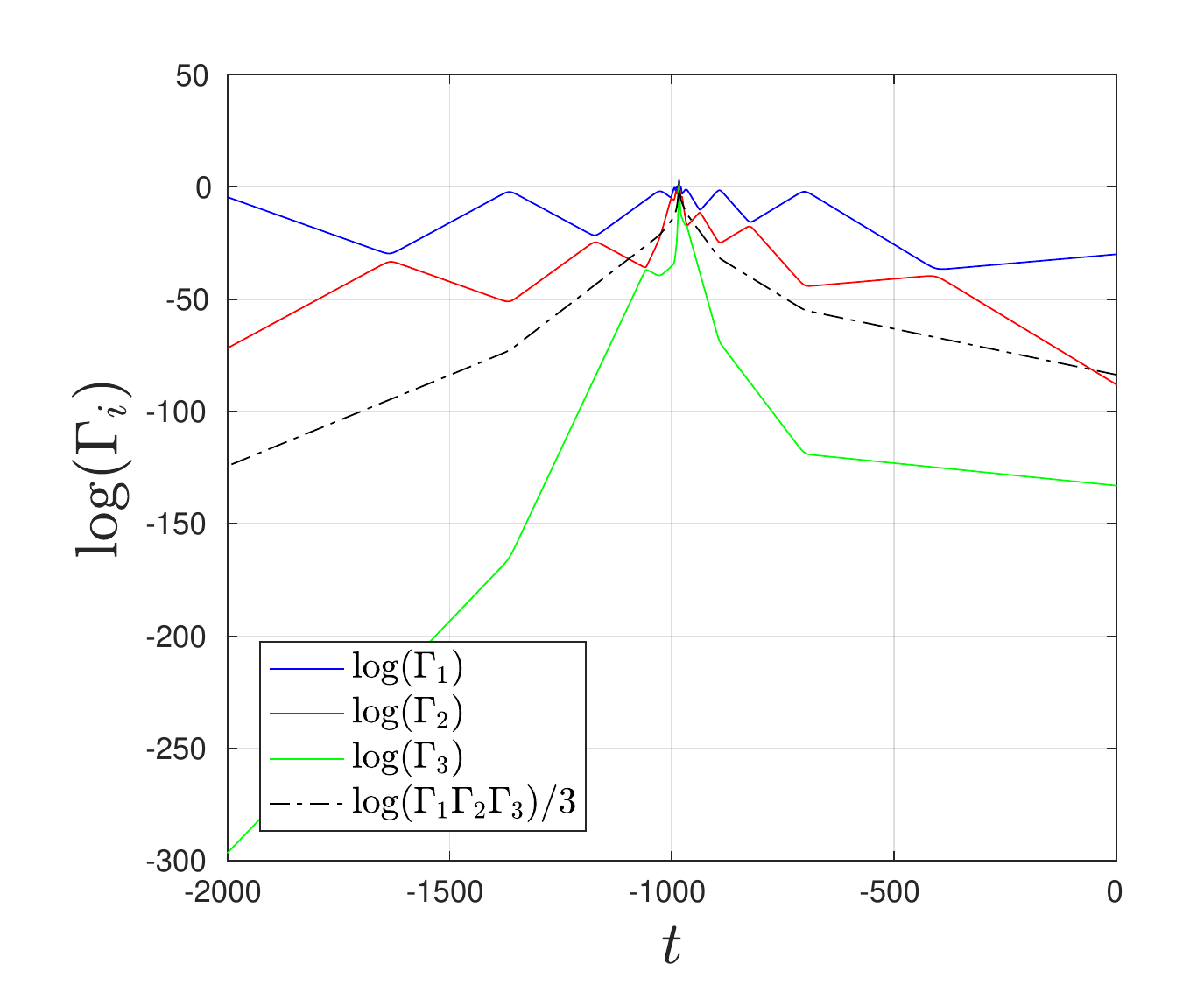}
		 \subcaption{Plot of $\Gamma_i$ variables ``close'' to the recollapse.}
			\label{fig:bkl_rebounce}
	\end{subfigure}
	\qquad \quad
	\begin{subfigure}[t]{0.4\textwidth}
		\includegraphics[width=8cm]{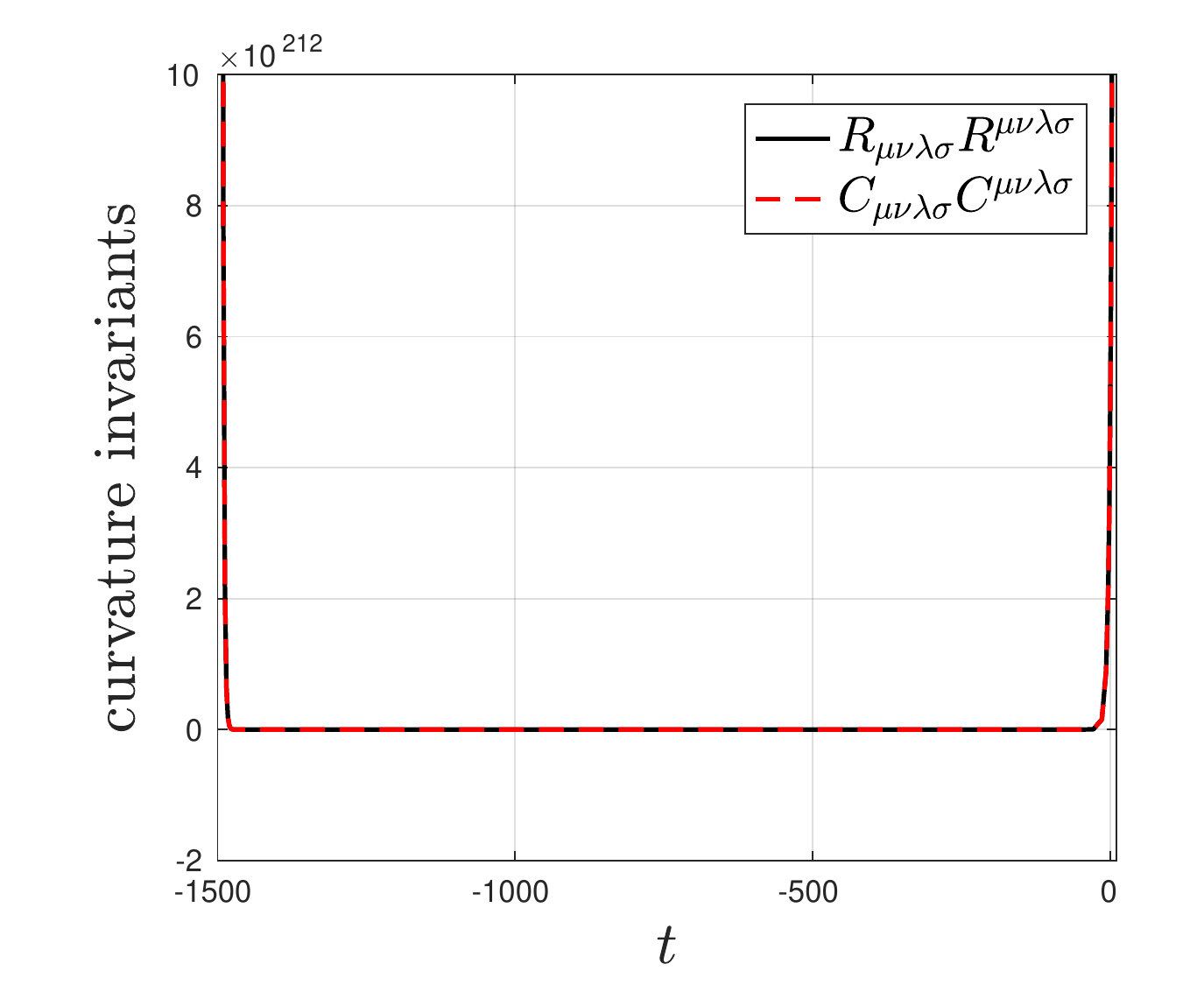}
			\subcaption{Plot of Kretschmann and Weyl squared scalar corresponding to FIG. \ref{fig:bkl_rebounce}.}
			\label{fig:Kretschmann_rebounce}
	\end{subfigure}
	\caption{Plot (a) shows the result of the application of the shooting method described in \cite{Kiefer:2018uyv} ``close'' to the recollapse (which is roughly at $t\approx -1000$). The plot (b) was obtained from a numerical calculation based on the result of section \ref{sec:Kretschmann}.}
	\label{fig:plots1}
\end{figure}

A numerical evaluation ``close'' to the point of recollapse\footnote{The existence of the recollapse was proven by Lin and Wald \cite{Wald}.}
is shown in Fig. \ref{fig:plots1}. We note that the bare Kretschmann scalar appears to roughly blow up exponentially in $t$ and it rapidly exceeds the range of numbers that are accessible in Matlab. This is why from now on we turn to numerically evaluating the so-called Hubble normalized Kretschmann scalar $R_{\mu\nu\lambda\sigma}R^{\mu\nu\lambda\sigma}/|K^i{}_i|$. This quantity has the virtue of being dimensionless and numerically well behaved. The expansion scalar is given by
\begin{equation}
K^i{}_i=\frac{1}{2N}\left[
(\log \Gamma_1)^\cdot+(\log \Gamma_2)^\cdot+(\log \Gamma_3)^\cdot \right] = \frac{3}{N}\dot{\alpha}
\ .
\end{equation}

\section{The asymptotic regime close to the singularity}
\label{sec:asymptotic_regime}

We remark again that we are considering the tumbling case. We expect,  however, that a similar discussion also holds for the non-tumbling and non-rotating cases.

In order to simplify the dynamics of Bianchi IX   (see, e.g. the review in \cite{bkl}) we make two assumptions.
The first assumption states that the anisotropy of space grows without
bound. This means that the solution enters the regime
	\begin{equation}
		\Gamma_1\gg \Gamma_2\gg \Gamma_3 \ ,
		\label{eq:BIX_Gamma_inequality}
	\end{equation}
where	the ordering of indices depends on the initial conditions and is mostly irrelevant for our discussion.
The second assumption  states that the Euler angles assume
constant values:
	\begin{equation}
	(\theta,\phi,\psi)\rightarrow (\theta_0,\phi_0,\psi_0) \ ,
	\end{equation}
that is, the rotation of the principal axes stops for all practical purposes and the metric becomes {\it effectively} diagonal. For our variables this means that the dust velocities $v_i$ assume constant values.
The main purpose of the article \cite{Kiefer:2018uyv} was to provide a numerical verification for the two assumptions.

According to the phrase ``matter does not matter''  we expect the matter terms in the Kretschmann scalar to be negligible in the asymptotic regime, that is, the Weyl part should dominate  over the Ricci part.

During Kasner epochs (i.e. between two successive bounces in the asymptotic regime) we expect the most relevant term   to be the term in the first two lines of \eqref{eq:E_squared_term1} right after the equality-sign.  We therefore assume now that the Kretschmann scalar can be approximated by
\begin{equation}
\begin{aligned}
R_{\mu\nu\lambda\sigma}R^{\mu\nu\lambda\sigma} \approx \frac{1}{N^4}
&
\left(
\left[ (\log \Gamma_1)^\cdot(\log \Gamma_2)^\cdot\right]^2
+\left[ (\log \Gamma_1)^\cdot(\log \Gamma_3)^\cdot\right]^2
+\left[ (\log \Gamma_2)^\cdot(\log \Gamma_3)^\cdot\right]^2
\right.
\\
&
\left. \
+(\log \Gamma_1)^\cdot(\log \Gamma_2)^\cdot(\log \Gamma_3)^\cdot
\left[(\log \Gamma_1)^\cdot + (\log \Gamma_2)^\cdot + (\log \Gamma_3)^\cdot \right]\right) \ .
\end{aligned}
\label{eq:Kasner_assumption}
\end{equation}
This claim is confirmed by our numerical simulations.
We remark that this term corresponds exactly to the Weyl squared scalar of the Bianchi I model
with the metric \begin{equation}
\dd s^2= -N^2 \dd t^2 + \Gamma_1 \dd x^2 + \Gamma_2 \dd y^2 + \Gamma_1 \dd z^2 \ .
\end{equation}
The Weyl tensor of the Bianchi I model has  only an electric part, and the magnetic part vanishes (in the quasi-Gaussian gauge).

\begin{figure}[!ht]
\hspace{-2.5cm}
	\begin{subfigure}[t]{0.4\textwidth}
		\includegraphics[width=8cm]{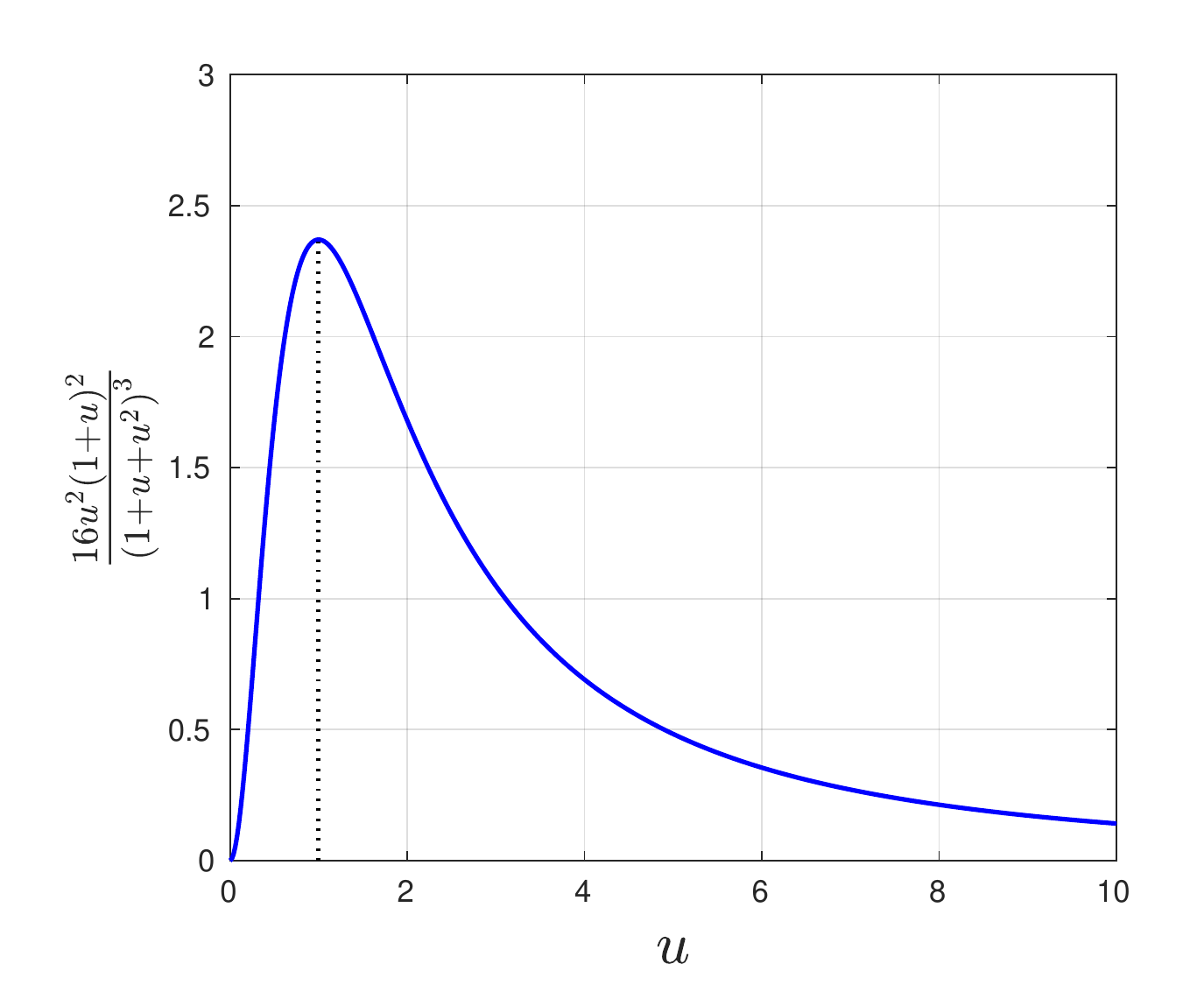}
		 \subcaption{Plot of the function $\frac{16 u^2 (1+u)^2}
			{\left(1+u+u^2\right)^3}$ on the right hand side of 	
			\eqref{eq:Kretschmann_u}.}
			\label{fig:Kretschmann_u}
	\end{subfigure}
	\qquad \quad
	\begin{subfigure}[t]{0.4\textwidth}
		\includegraphics[width=8cm]{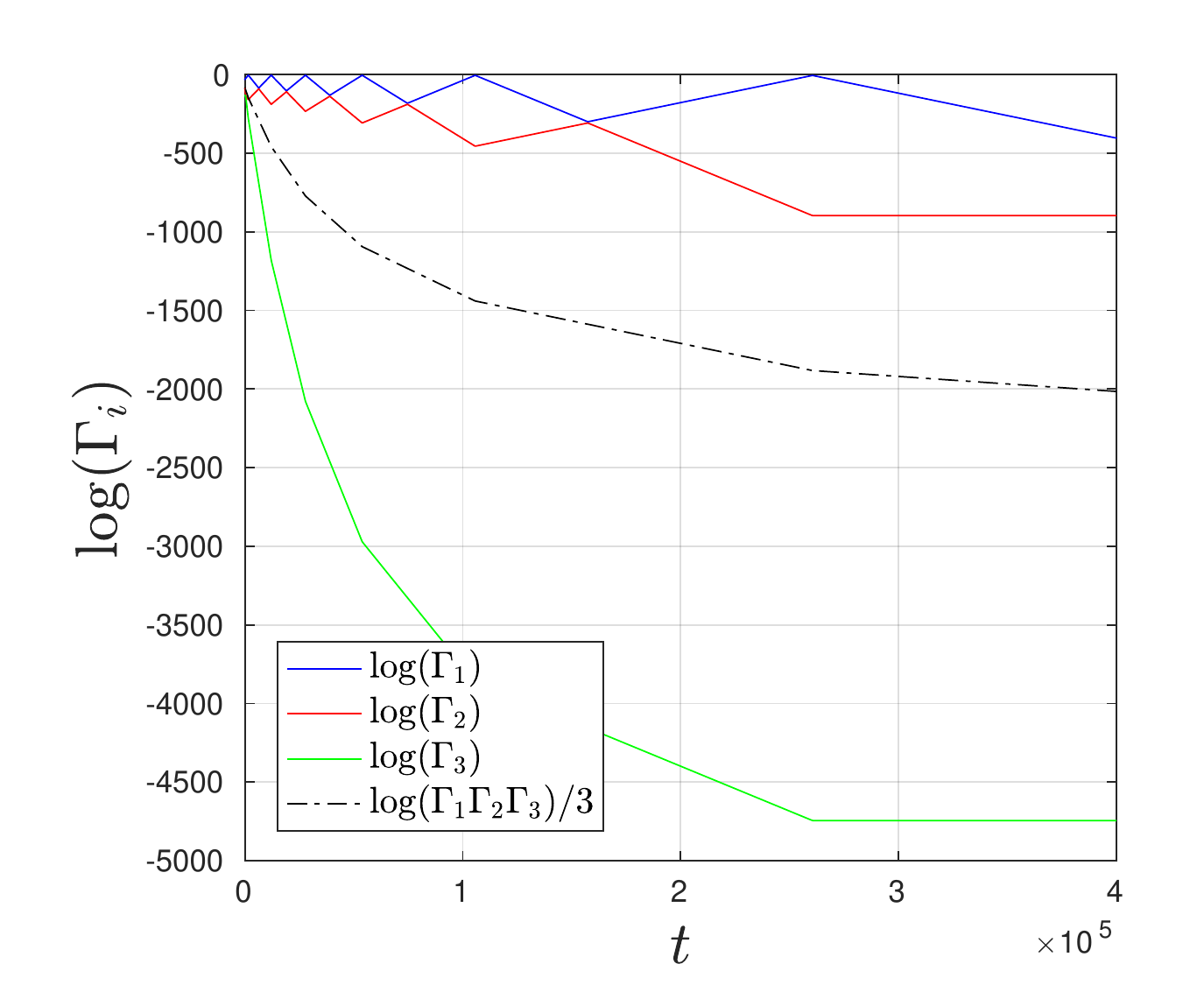}
			\subcaption{Plot of the variables $\log (\Gamma_i)$ obtained via the shooting method.}
			\label{fig:bkl}
	\end{subfigure}\\
\hspace{-2.5cm}
	\begin{subfigure}[t]{0.4\textwidth}
		\includegraphics[width=8cm]{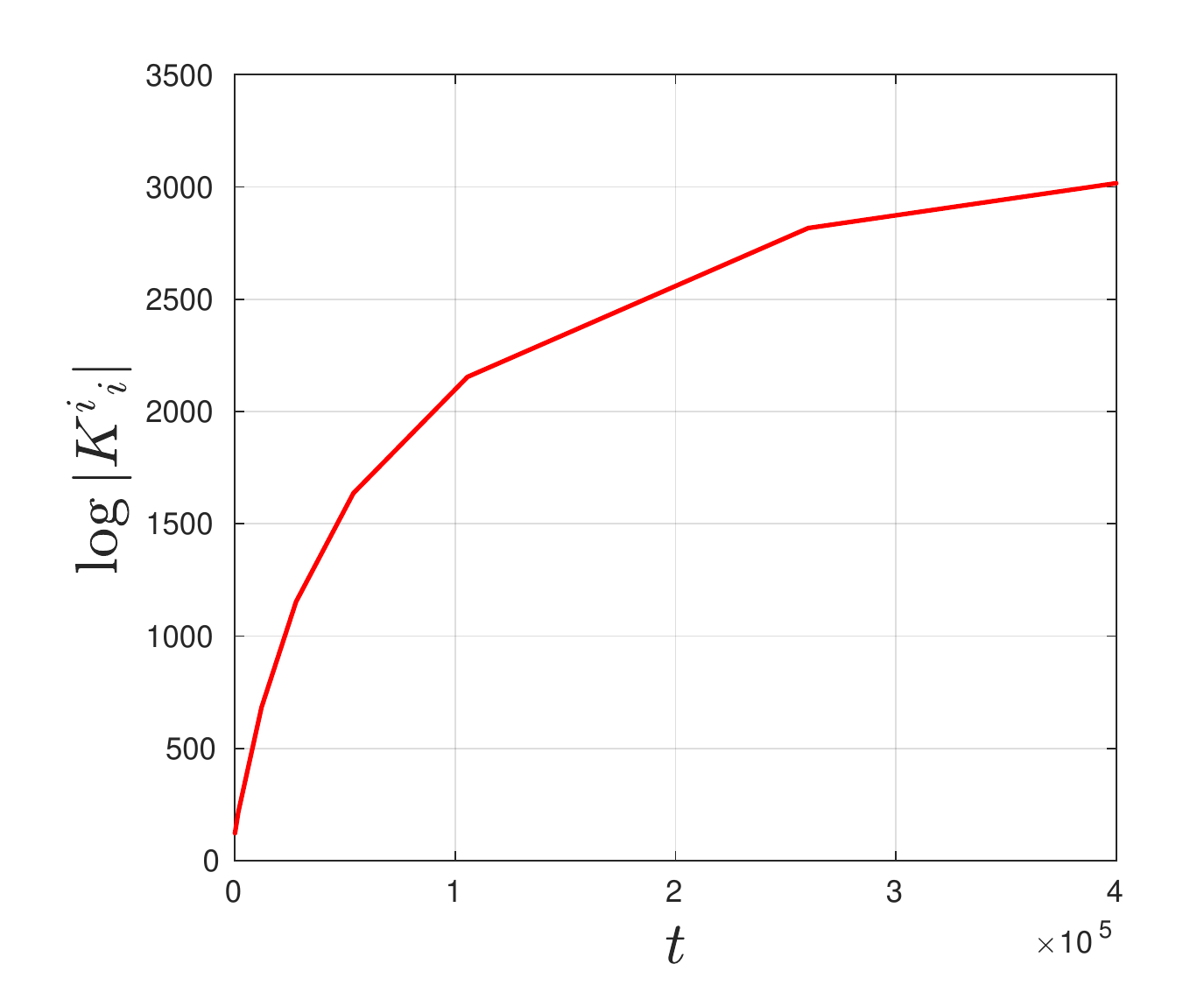}
		\subcaption{Plot of the logarithm of the expansion.}
			\label{fig:expansion}
	\end{subfigure}
	\qquad \quad
	\begin{subfigure}[t]{0.4\textwidth}
		\includegraphics[width=8cm]{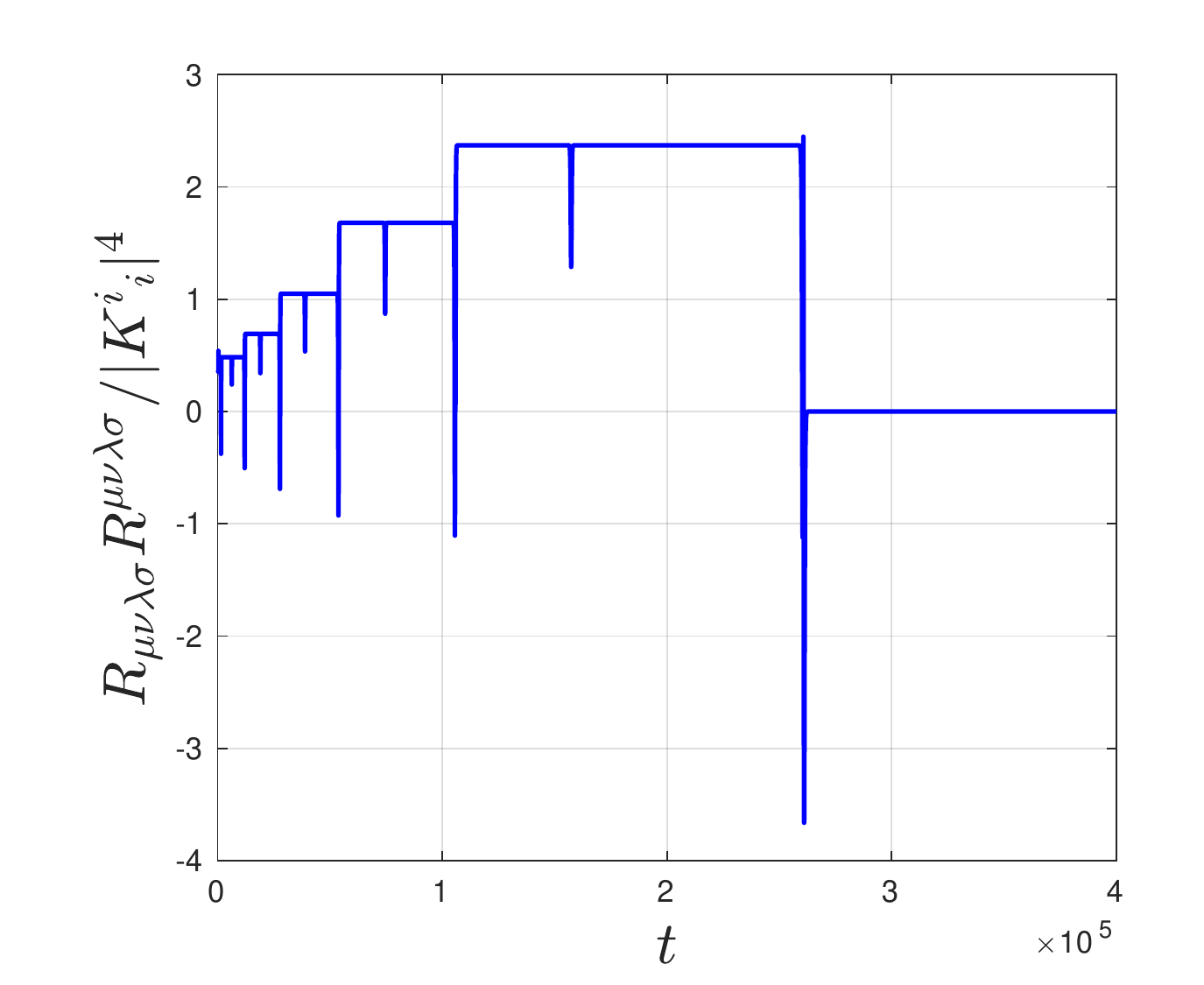}
		\subcaption{Plot of the Hubble normalized Kretschmann scalar. }
		\label{fig:Weyl squared}
	\end{subfigure}
	\caption{The plot (b) was obtained by the shooting method described in \cite{Kiefer:2018uyv}. One can  see a typical Kasner era composed of epochs that are approximately connected by transformations of the first and second kind. The output of the shooting method was used to obtain the plots (c) and (d) based on the calculation in section \ref{sec:Kretschmann} of this paper.  }
	\label{fig:plots2}
\end{figure}

During Kasner epochs the time evolution  can be parameterized using the Lifshitz -Khalatnikov parameter $u$ following the considerations in \cite{bkl}.
Doing so and using the assumption \eqref{eq:Kasner_assumption} we obtain that the Hubble-normalized Kretschmann scalar
 can be approximated by
\begin{equation}
R_{\mu\nu\lambda\sigma}R^{\mu\nu\lambda\sigma}/|K^i{}_i|^4 \approx\frac{16 u^2 (1+u)^2}
{\left(1+u+u^2\right)^3} \qquad \text{during Kasner epochs} .
\label{eq:Kretschmann_u}
\end{equation}
Consequently the Kretschmann scalar blows up like the expansion $K^i{}_i$ to the power $4$ during Kasner eras.
In order to understand the temporal evolution of the Kretschmann scalar over the course of one epoch we have plotted the expression on the right hand side of \eqref{eq:Kretschmann_u} as a function of $u$ in Fig. \ref{fig:Kretschmann_u}. It is   important that the function has a maximum in $u=1$. \footnote{This maximum implies an upper bound for the Hubble normalized Kretschmann scalar during Kasner eras given by $64/27.$}

 The BKL refers to bounces from the curvature potential as {\em transformations  of the first kind} while they call bounces from  centrifugal walls {\em transformations of the second kind}.
Transformations of the first kind change   the Lifshitz-Khalatnikov parameter according to
$
u\overset{1}{\rightarrow} u-1 $. Transformations of the second kind interchange the values of the velocities according to $(\log \Gamma_1)^\cdot\overset{2}{\rightarrow} (\log \Gamma_2)^\cdot$, $(\log \Gamma_2)^\cdot\overset{2}{\rightarrow} (\log \Gamma_1)^\cdot$ and leave the value of $u$  unchanged, i.e. $u \overset{2}{\rightarrow}u$.
 It follows that $\overset{1}{\rightarrow}$ changes the value of the Hubble normalized Kretschmann scalar \eqref{eq:Kretschmann_u} while $\overset{2}{\rightarrow}$ does not.
 According to the analysis in \cite{bkl}  a typical Kasner era can be expressed as a sequence of $n$ Kasner epochs which starts  with an epoch that has a maximum $u$-value larger than $1$ when evolving towards the singularity. The value of $u$ decreases with each transformation of the first kind  and ends with the epoch for which $u$ becomes  smaller than $1$ for the first time, e.g.
\begin{equation}
1<u_1=u_{\text{max}}
\overset{1}{\rightarrow}
u_2
\overset{2}{\rightarrow}
u_3
\overset{1}{\rightarrow}
u_4
\overset{2}{\rightarrow}
u_5
\overset{1}{\rightarrow}
\ \ldots \
\overset{2}{\rightarrow}u_{n-1}\overset{1}{\rightarrow} u_n=u_{\text{min}}<1 \ .
\label{eq:u_map}
\end{equation}
It should be  remarked at this point that the $u$-map was found to be asymptotically exact for particular cases (for a collection of rigorous results concerning the $u$-map see \cite{Uggla}).
 A solution of the discrete mixmaster map  and  a detailed study of its chaotic nature for the vacuum Bianchi IX case can be found in \cite{Chernoff_Barrow_1972}.

We are now in the position to provide a picture of the behaviour of the Kretschmann scalar over the course of one Kasner era:
According  to the formula  \eqref{eq:Kretschmann_u} plotted in  Fig. \ref{fig:Kretschmann_u} and \eqref{eq:u_map} we  expect the Hubble normalized Kretschmann scalar to increase its value with each transformation of the first kind before it hits the value $u_{\text{min}}<1$ for which it decreases again.
 This is apart from the behaviour in the vicinity of the bounces precisely what we observe in the numerically evaluated Hubble normalized scalar plotted in Fig. \ref{fig:plots2}. We remark that we regard the part of the solution plotted in Fig. \ref{fig:bkl} to be not quite in the asymptotic regime but rather at the tansition into the asymptotic regime.

It was helpful, in this paper,  to support the computations by using the tensor algebra package
{\it xAct} \cite{xAct}. Numerical calculations were performed using MATLAB R2016b.

\section{Summary}
\label{con}

The main purpose of this paper is to provide  a description of the temporal behaviour of the Kretschmann scalar in the asymptotic regime.
 In this regime the volume density, being proportional to the product of the three directional scale factors, evolves towards zero \cite{bkl},
but this is not a satisfactory indication of the singularity as it depends on the choice of coordinates.
The blowing up of
curvature invariants, on the other hand,
shows that we are dealing with a genuine  curvature singularity.

During Kasner epochs, $R_{\mu\nu\lambda\sigma}R^{\mu\nu\lambda\sigma}$ increases like the expansion to the power four.
Over the course of a single Kasner era  the value of the Hubble normalized Kretschmann scalar increases until it  drops down to a finite
value when it ends. This process will repeat itself with the beginning of the next Kasner era  until the system approaches the singularity.

The present paper is supposed to be
an extension  of our previous paper \cite{Kiefer:2018uyv}, which considers, for simplicity, only the  tilted dust field as a source. The discussion of other tilted fluids goes beyond
the scope of our present programme.
The effect of  tilted radiation, which has been studied analytically and numerically in the recent paper \cite{Ganguly:2017qff}, is  particularly interesting.


The asymptotic regimes of the Bianchi IX and BVIII models are quite similar  \cite{VBiel}.
Both models have been used to derive the BKL scenario \cite{BKL22}.

The asymptotic regime approximates well the dynamics near the singularity.
This is why it was recently used in the struggle for removing the singularity  of the BKL scenario  by quantization \cite{AWG}. \\

\acknowledgments This paper profited from correspondence with Vladimir Belinski and Claes Uggla. Moreover we would like to thank Claus Kiefer
for helpful discussions.   We are also grateful to the anonymous referee for constructive criticism. This work was supported by the German-Polish
bilateral project DAAD and MNiSW, No 57391638.

\newcommand{\journal}{\rm}
\newcommand{\booktitle}{\em}
\newcommand{\vol}{\bf}
\newcommand{\papertitle}{\em}
\newcommand{\publisher}{\rm}

\end{document}